\documentclass[aip,apl,reprint]{revtex4}
\usepackage{amsmath}

\usepackage{graphicx}
\usepackage{dcolumn}
\usepackage{bm}
\usepackage[mathlines]{lineno}

\begin{document}

\title{Graphene induced mode bifurcation at low input power}
\author{Rujiang Li}
\author{Xiao Lin}
\affiliation{State Key Laboratory of Modern Optical Instrumentation,
Zhejiang University, \\Hangzhou 310027, China}\affiliation{College of
Information Science and Electronic Engineering, Zhejiang University,
\\Hangzhou 310027, China}\affiliation{The Electromagnetics Academy of
Zhejiang University, Zhejiang University, \\Hangzhou 310027, China}
\author{Shisheng Lin}
\affiliation{State Key Laboratory of Modern Optical Instrumentation,
Zhejiang University, \\Hangzhou 310027, China}\affiliation{College of
Information Science and Electronic Engineering, Zhejiang University,
\\Hangzhou 310027, China}
\author{Xianmin Zhang}
\author{Erping Li}
\affiliation{College of Information Science and Electronic Engineering,
Zhejiang University, \\Hangzhou 310027, China}
\author{Hongsheng Chen}
\thanks{To whom correspondence should be addressed:
E-mail: hansomchen@zju.edu.cn (H. Chen); liep@zju.edu.cn (E. Li).}
\affiliation{State Key Laboratory of Modern Optical Instrumentation,
Zhejiang University, \\Hangzhou 310027, China}\affiliation{College of
Information Science and Electronic Engineering, Zhejiang University,
\\Hangzhou 310027, China}\affiliation{The Electromagnetics Academy of
Zhejiang University, Zhejiang University, \\Hangzhou 310027, China}
\date{\today}

\begin{abstract}
We study analytically the plasmonic modes in the graphene-coated dielectric nanowire,
based on the explicit form of nonlinear surface
conductivity of graphene. The propagation constants of different plasmonic modes can
be tuned by the input power at the order of a few tenths of mW. The
lower and upper mode bifurcation branches are connected at the limitation
value of the input power. Moreover, due to the nonlinearity of graphene, the
dispersion curves of plasmonic modes at different input powers form
an energy band, which is in sharp contrast with the single dispersion curve
in the limit of zero input power.
\end{abstract}

\maketitle

Nonlinear plasmonics is a newly developed but explosive growing field which
not only offers extreme light manipulation at the subwavelength scale \cite%
{nature424-824}, but also provides an universal method to scale down the
conventional nonlinear optical devices to the chip scale \cite{nphoton6-737}%
. Due to the strong local electromagnetic fields, nonlinear plasmonic
effects can originate from the adjacent nonlinear dielectric media. In the
past years, nonlinear plasmonic modes in metal-dielectric\cite%
{SPJETPL32-512,PRA28-1855,OE17-21732,PRA81-033850},
dielectric-metal-dielectric\cite%
{JOSA72-1345,OL9-235,JAP58-2460,OL32-674,OE19-6616}, and
metal-dielectric-metal \cite{OE16-21209,PRA91-043815} planar structures have been
studied extensively. Meanwhile, the existence of discrete solitons in
nonlinear dielectric media embedded with periodic metallic films \cite%
{PRL99-153901}, nanowires \cite{PRL104-106803}, and nanorings \cite%
{OE20-1945} have been proposed recently.

As a newly discovered two dimensional electromagnetic material, graphene
has received extensive attention in nonlinear plasmonics.
Compared with ordinary dielectric media, graphene has a high
nonlinear susceptibility
\cite{EPL79-27002,JP20-384204,PRL105-097401,PRB82-201402,PR535-101},
which is promising to release the demand of high
input power in current nonlinear plasmonics. Until now, some basic
phenomena based on graphene induced nonlinearities have been
studied, e.g., solitons supported by monolayer graphene \cite{LPR7-L7} and
multilayer graphene \cite{JPB46-155401,LPR8-291,PRB91-045424}. However,
little attention has been paid to the two dimensional graphene-based
nonlinear plasmonic waveguides, although two dimensional structures are
more favourable as fundamental building blocks of plasmonic waveguide arrays
and plasmonic lattices \cite{PRL104-106803,OE20-1945}.

In this Letter, we give explicitly the tensor form of nonlinear surface
conductivity of graphene in the classical frequency range. As a simplest
structure of the two dimensional waveguides, the plasmonic modes with
different orders are presented analytically in the graphene-coated
dielectric nanowire, where the nonlinearity of graphene is considered.
Meanwhile, the dependencies between the propagation constants and the input power
are discussed. Moreover, the energy band which is formed by dispersion cures at
different input powers, is studied by taking the fundamental mode as an example.

Due to the two dimensional nature of graphene, its third order surface
conductivity is a fourth-order tensor with $16$ elements. Previous studies
mainly focus on the element $\sigma_{xxxx}^{(3)}$ \cite%
{EPL79-27002,JP20-384204,PRB90-125425}, whereas
other elements are needed when the incident electric field has two
components along the graphene surface. For this reason, we assume that the
doped graphene monolayer is placed on the $xy$ plane and a time-dependent
electric field of the form $\mathbf{E}\left( t\right) =\left[
E_{x}\exp\left( -i\omega t\right) +c.c.\right] \hat{x}+\left[
E_{y}\exp\left( -i\omega t\right) +c.c.\right] \hat{y}$ is applied, where $%
\omega$ is the angular frequency of the electromagnetic field. Basically the
optical response of graphene is contributed by both the intraband and
interband electronic transitions \cite{PRB82-201402,PR535-101}. However, in the
classical frequency range $\hbar\omega\leq\mu_{c}$, namely the photon energy
is smaller than the chemical potential, the optical response of graphene is
dominated by the intraband transitions. Under the relaxation time
approximation and neglecting the interband processes, the transport
properties of electrons in graphene are governed by the following Boltzmann
equation{
\begin{equation}
\frac{\partial f\left( \mathbf{k},t\right) }{\partial t}-\frac{e}{\hbar }%
\mathbf{E}\left( t\right) \mathbf{\cdot}\frac{\partial f\left( \mathbf{k}%
,t\right) }{\partial\mathbf{k}}=-\frac{f\left( \mathbf{k},t\right)
-f_{0}\left( \mathbf{k}\right) }{\tau},  \label{Boltzmann}
\end{equation}
}where $\mathbf{k=}\left( k_{x},k_{y}\right) $ is the wave vector, $f\left(
\mathbf{k},t\right) \ $ is the nonequilibrium distribution function, $%
f_{0}\left( \mathbf{k}\right) =1/\left[ 1+e^{\left( \epsilon\left( \mathbf{k}%
\right) -\mu_{c}\right) /k_{B}T}\right] $ is the equilibrium Fermi-Dirac
distribution function, $\epsilon\left( \mathbf{k}\right) =v_{F}\hbar\sqrt{%
k_{x}^{2}+k_{y}^{2}}$ is the Dirac cone spectrum of charge carriers in
graphene, $v_{F}=c/300$ is the Fermi velocity, $c$ is the velocity of light
in free space, $\tau$ is the relaxation time, $-e$ is the charge of an
electron, $\hbar$ is the reduced Plank's constant, $k_{B}$ is the
Boltzmann's constant, and $T$ is the temperature.

The exact solution of Eq. (\ref{Boltzmann}) at $\omega\tau\gg1$
is \cite{PRB90-125425,PRB91-045424} {
\begin{equation}
f\left( \mathbf{k},t\right) =\frac{e^{-t/\tau}}{\tau}\int_{-\infty}^{t}dt^{%
\prime}e^{t^{\prime}/\tau}f_{0}\left[ \mathbf{k}+{\kappa}\left(
t,t^{\prime}\right) \right] ,  \label{solution_Boltzmann}
\end{equation}
}where ${\kappa}\left( t,t^{\prime}\right) =\frac{e}{\hbar}%
\int_{t^{\prime}}^{t}\mathbf{E}\left( t^{\prime\prime}\right)
dt^{\prime\prime}$.
The surface current along the graphene surface can be expressed as{
\begin{equation}
\mathbf{j}\left( t\right) =\mathbf{-}4\frac{e}{\left( 2\pi \right) ^{2}\hbar
}\int d\mathbf{k}f\left( \mathbf{k},t\right) \frac{\partial \epsilon \left(
\mathbf{k}\right) }{\partial \mathbf{k}},  \label{current}
\end{equation}%
}where the factor $4$ is due to spin degeneracy and valley degeneracy \cite%
{PRB91-045424}. In the low temperature limit $T\rightarrow 0$, the
nonequilibrium distribution function can be replaced by the Heaviside step
function.
Thus the surface current reduces to {
\begin{equation}
j_{i}\left( t\right) \mathbf{=}\mathbf{-}\frac{ev_{F}}{\pi ^{2}\tau }%
e^{-t/\tau }\int_{-\infty }^{t}dt^{\prime }e^{t^{\prime }/\tau }I_{i}\left(
t,t^{\prime }\right) ,  \label{current_i}
\end{equation}%
}where{
\begin{equation}
I_{i}\left( t,t^{\prime }\right) =\int d\mathbf{k}\frac{k_{i}}{\sqrt{%
k_{x}^{2}+k_{y}^{2}}}H\left[ \mu _{c}-\epsilon \left( \mathbf{k}+{\kappa }%
\left( t,t^{\prime }\right) \right) \right] ,  \label{integral}
\end{equation}%
}and $i=x,y$. Eq. (\ref{integral}) can be calculated by expanding the
integrand with respect to $\kappa _{x}\left( t,t^{\prime }\right) $ and $%
\kappa _{y}\left( t,t^{\prime }\right) $ (up to the third order) and
integrating over the Fermi surface. Calculation shows that {
\begin{equation}
I_{i}\left( t,t^{\prime }\right) =-k_{F}\pi \kappa _{i}\left( t,t^{\prime
}\right) +\frac{\pi }{8k_{F}}\kappa _{i}\left( t,t^{\prime }\right) \kappa
^{2}\left( t,t^{\prime }\right) ,  \label{integral_i}
\end{equation}%
}where $k_{F}=\mu _{c}/$$v_{F}\hbar $ is the Fermi wave vector. Without loss
of generality, we consider the surface current in $x$ direction. According
to Eqs. (\ref{current_i}) and (\ref{integral_i}), and considering the
equivalence between $x$ and $y$ coordinates, we obtain {
\begin{equation}
j_{x}=j_{x}\left( \omega \right) \exp \left( -i\omega t\right) +j_{x}\left(
3\omega \right) \exp \left( -i3\omega t\right) +c.c,  \label{current_x}
\end{equation}%
}where the first term corresponds to the surface current with time
dependence $\exp \left( -i\omega t\right) $ and the second term corresponds
to the third harmonics. In what follows, we neglect the terms of third
harmonics since the phase matching condition is required\cite{NO}, and
obtain {
\begin{eqnarray}
j_{x}\left( \omega \right)  &=&\sigma _{xx}^{\left( 1\right) }E_{x}+[\sigma
_{xxyy}^{\left( 3,\omega \right) }+\sigma _{xyyx}^{\left( 3,\omega \right)
}]E_{x}\left\vert E_{y}\right\vert ^{2}+\sigma _{xyxy}^{\left( 3,\omega
\right) }E_{x}^{\ast }E_{y}^{2}  \nonumber \\
&&+\sigma _{xxxx}^{\left( 3,\omega \right) }\left\vert E_{x}\right\vert
^{2}E_{x}.  \label{current_x_omeg}
\end{eqnarray}%
}Thus $\sigma _{xx}^{(1)}$ $=\sigma _{yy}^{\left( 1\right) }=4\sigma _{0}\mu
_{c}/\pi \hbar (1/\tau -i\omega )$ is the linear surface conductivity, $%
\sigma _{xxyy}^{\left( 3,\omega \right) }=\sigma _{xyxy}^{\left( 3,\omega
\right) }=\sigma _{xyyx}^{\left( 3,\omega \right) }=\sigma _{yyxx}^{\left(
3,\omega \right) }=\sigma _{yxyx}^{\left( 3,\omega \right) }=\sigma
_{yxxy}^{\left( 3,\omega \right) }=\sigma _{xxxx}^{\left( 3,\omega \right)
}/3=\sigma _{yyyy}^{\left( 3,\omega \right) }/3=-3\sigma
_{0}e^{2}v_{F}^{2}/\pi \mu _{c}\hbar (1/\tau ^{2}+\omega ^{2})(1/\tau
-i2\omega )$ are the nonzero elements of nonlinear surface conductivity, and
$\sigma _{0}=e^{2}/4\hbar $ is the conductivity quantum.

Under the limit of $\omega\tau\gg1$, namely the relaxation time is large
compared with the oscillation period of the incident electromagnetic field, the
surface conductivity of graphene is{\small
\begin{equation}
\sigma=\sigma^{\left( 1\right) }+\sigma^{\left( 3\right) }\left\vert \mathbf{%
E}_{\parallel}\right\vert ^{2},  \label{sigma}
\end{equation}
}where{\small
\begin{equation}
\sigma^{\left( 1\right) }=i\frac{e^{2}\mu_{c}}{\pi\hbar^{2}\omega}
\label{sigma1}
\end{equation}
}is the linear part of surface conductivity,{\small
\begin{equation}
\sigma^{\left( 3\right) }=-i\frac{9e^{4}v_{F}^{2}}{8\pi\mu_{c}\hbar
^{2}\omega^{3}}  \label{sigma3}
\end{equation}
}is the nonlinear part of surface conductivity, and $\mathbf{E}_{\parallel}$
is the electric field that is parallel to the graphene surface. Note in
deriving Eqs. (\ref{sigma})-(\ref{sigma3}), we have used the condition $%
E_{y}^{\ast}E_{x}=E_{x}^{\ast}E_{y}$, which requires that the electric field
$\mathbf{E}_{\parallel}$ is linearly polarized. If the incident wave is
circularly polarized or elliptically polarized, the tensor form of nonlinear
surface conductivity should be used.

Besides, Eqs. (\ref{sigma})-(\ref{sigma3})
are valid under the approximation conditions of $\hbar\omega\leq\mu_{c}$ and
$\omega\tau\gg1$. The carrier relaxation time $\tau$ is determined by the carrier
mobility $\mu$ as $\tau=\mu\mu_{c}/ev_{F}^{2}$, where the carrier mobility $\mu$ of
graphene film ranges from 1000 cm$^{2}$/(V$\cdot$ s) in chemical vapor deposition
(CVD)-grown graphene \cite{nnano6-630} to 230 000 cm$^{2}$/(V$\cdot$ s) in suspended exfoliated graphene
\cite{SSC146-351}. When using a moderate mobility of 10 000 cm$^{2}$/(V$\cdot$ s)
and with $\mu_{c}=0.3$ eV, the frequency of the electric field satisfies
$0.5$ THz $\ll \omega /2\pi \leq 45.3$ THz.

\begin{figure}[ptb]
\centering
\vspace{-0.1cm} \includegraphics[width=5.5cm]{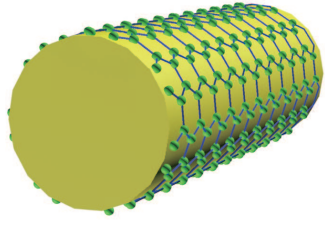} \vspace{-0.2cm}
\caption{Structure of the graphene-coated dielectric nanowire.}
\label{structure}
\end{figure}

Considering the practical applications where two dimensional structures are
more favourable, in the following we calculate the plasmonic modes of
graphene-coated dielectric nanowire, where the structure
is shown in Fig. \ref{structure}.
For the modes propagating in the axial direction of the waveguide,
the field can be expressed as
{
\begin{align}
\mathbf{\tilde{A}}\left( r,\theta ,z,t\right) & =\mathbf{A}\left( r,\theta
\right) \exp \left[ i\left( \beta z-\omega t\right) \right]+c.c. ,
\label{E_expression}
\end{align}%
}where $\mathbf{\tilde{A}}$ denotes the electric field $\mathbf{\tilde{E}}$
or the magnetic field $\mathbf{\tilde{H}}$, $\beta $ is the
propagation constant, and $z$ is the propagation direction. From Eq. (%
\ref{E_expression}), the $z$ component of the field satisfies the following
equation {
\begin{align}
& \frac{\partial ^{2}A_{z}}{\partial r^{2}}+\frac{1}{r}\frac{\partial A_{z}}{%
\partial r}+\frac{1}{r^{2}}\frac{\partial ^{2}A_{z}}{\partial \theta ^{2}}%
-\left( \beta ^{2}-k_{0}^{2}\varepsilon \right) A_{z}=0,  \label{Ez_equation}
\end{align}%
}where $k_{0}=\omega \sqrt{\varepsilon _{0}\mu _{0}}$, and $\varepsilon $ is
the relative permittivity of the material ($\varepsilon =\varepsilon _{1}$
for the inside dielectric nanowire and $\varepsilon =\varepsilon _{2}=1$ for
the outside air). Using the method of separating variables, the solutions
for the $m$-th order plasmonic mode are {
\begin{align}
E_{z}\left( r,\theta \right) & =i\frac{\sqrt{P_{0}\eta _{0}}}{a}%
A_{m}I_{m}\left( u\frac{r}{a}\right) e^{im\theta },  \label{1} \\
H_{z}\left( r,\theta \right) & =\frac{\sqrt{P_{0}/\eta _{0}}}{a}%
B_{m}I_{m}\left( u\frac{r}{a}\right) e^{im\theta },  \label{2}
\end{align}%
}for $r\leq a$, and {
\begin{align}
& E_{z}\left( r,\theta \right) =i\frac{\sqrt{P_{0}\eta _{0}}}{a}%
C_{m}K_{m}\left( w\frac{r}{a}\right) e^{im\theta },  \label{3} \\
& H_{z}\left( r,\theta \right) =\frac{\sqrt{P_{0}/\eta _{0}}}{a}%
D_{m}K_{m}\left( w\frac{r}{a}\right) e^{im\theta },  \label{4}
\end{align}%
}for $r>a$, where $a$ is the radius of the dielectric nanowire,
$u=a\sqrt{\beta ^{2}-k_{0}^{2}\varepsilon _{1}}$, $w=a%
\sqrt{\beta ^{2}-k_{0}^{2}\varepsilon _{2}}$, $A_{m}$, $B_{m}$, $C_{m}$ and $%
D_{m}$ are the dimensionless undetermined constants, $P_{0}=\frac{1}{4}%
\iint_{S}\left( \mathbf{E\times H}^{\ast }+\mathbf{E}^{\ast }\times \mathbf{H%
}\right) \cdot \hat{z}dS$ is the input power, $\eta _{0}=\sqrt{\mu
_{0}/\varepsilon _{0}}$ is the impendence of free space, and $I_{m}$ and $%
K_{m}$ are the $m$-th order modified Bessel function of the first kind and
the second kind, respectively. Utilizing the above results, the other
components of the electric field and magnetic field can be obtained from Maxwell
equations \cite{OE22-24322}. Since the electric field that is parallel to
the graphene surface is linearly polarized which satisfies $E_{y}^{\ast
}E_{x}=E_{x}^{\ast }E_{y}$, we can use Eqs. (\ref{sigma})-(\ref{sigma3}) to
characterize the surface conductivity of graphene directly. Besides, since
there are two components along the graphene surface, the contributions from
both two components should been considered, which is different to the case
of planar structures \cite{LPR7-L7,LPR8-291,PRB91-045424}. According to the
continuity conditions at $r=a$ and at arbitrary $\theta $, we can get the
following equations{
\begin{align}
& A_{m}I_{m}\left( u\right) =C_{m}K_{m}\left( w\right) ,  \label{21} \\
& B_{m}I_{m}\left( u\right) -D_{m}K_{m}\left( w\right)  \nonumber \\
=& i\frac{\sigma \eta _{0}a}{u^{2}}\left[ \beta mA_{m}I_{m}\left( u\right)
+k_{0}uB_{m}I_{m}^{\prime }\left( u\right) \right] ,  \label{22} \\
& \frac{1}{u^{2}}\left[ \beta mA_{m}I_{m}\left( u\right)
+k_{0}uB_{m}I_{m}^{\prime }\left( u\right) \right]  \nonumber \\
=& \frac{1}{w^{2}}\left[ \beta mC_{m}K_{m}\left( w\right)
+k_{0}wD_{m}K_{m}^{\prime }\left( w\right) \right] ,  \label{23} \\
& \frac{1}{u^{2}}\left[ k_{0}\varepsilon _{1}uA_{m}I_{m}^{\prime }\left(
u\right) +\beta mB_{m}I_{m}\left( u\right) \right]  \nonumber \\
-& \frac{1}{w^{2}}\left[ k_{0}\varepsilon _{2}wC_{m}K_{m}^{\prime }\left(
w\right) +\beta mD_{m}K_{m}\left( w\right) \right]  \nonumber \\
=& -i\frac{\sigma \eta _{0}}{a}A_{m}I_{m}\left( u\right) ,  \label{24}
\end{align}
}where {
\begin{align}
& \sigma =\sigma ^{\left( 1\right) }+\sigma ^{\left( 3\right) }\left\vert
\mathbf{E}_{\parallel }\right\vert ^{2}=\sigma ^{\left( 1\right) }+\sigma
^{\left( 3\right) }\left( \left\vert E_{z}\left( a\right) \right\vert
^{2}+\left\vert E_{\theta }\left( a\right) \right\vert ^{2}\right) ,
\label{25} \\
& E_{z}\left( a\right) =i\frac{\sqrt{P_{0}\eta _{0}}}{a}A_{m}I_{m}\left(
u\right) ,  \label{26} \\
& E_{\theta }\left( a\right) =i\frac{\sqrt{P_{0}\eta _{0}}}{u^{2}}[\beta
mA_{m}I_{m}\left( u\right) +k_{0}uB_{m}I_{m}^{\prime }\left( u\right) ],
\label{27}
\end{align}%
}and $m=0,1,\cdots $. Meanwhile, since the input power is $P_{0}$, we obtain
the normalization condition as follows{
\begin{gather}
\frac{\pi \beta k_{0}}{u^{4}}\left( \varepsilon
_{1}A_{m}^{2}+B_{m}^{2}\right) \int\nolimits_{0}^{a}\left( \frac{m^{2}a^{2}}{%
r^{2}}I_{m}^{2}+u^{2}I_{m}^{\prime 2}\right) rdr  \nonumber \\
+\frac{2\pi a}{u^{3}}A_{m}B_{m}\left( \varepsilon _{1}k_{0}^{2}+\beta
^{2}\right) \int\nolimits_{0}^{a}I_{m}I_{m}^{\prime }dr  \nonumber \\
+\frac{\pi \beta k_{0}}{w^{4}}\left( \varepsilon
_{2}C_{m}^{2}+D_{m}^{2}\right) \int\nolimits_{a}^{\infty }\left( \frac{%
m^{2}a^{2}}{r^{2}}K_{m}^{2}+u^{2}K_{m}^{\prime 2}\right) rdr  \nonumber \\
+\frac{2\pi a}{w^{3}}C_{m}D_{m}\left( \varepsilon _{2}k_{0}^{2}+\beta
^{2}\right) \int\nolimits_{a}^{\infty }K_{m}K_{m}^{\prime }dr=1.
\label{normalization}
\end{gather}%
}Thus, from the continuity conditions (\ref{21})-(\ref{27}) and the
normalization condition (\ref{normalization}), we can solve the five unknown
parameters $A_{m}$, $B_{m}$, $C_{m}$, $D_{m}$ and $\beta $ numerically.
Specially, when the nonlinear surface conductivity of graphene is neglected,
namely $\sigma ^{\left( 3\right) }=0$, our nonlinear plasmonic waveguide
reduces to the common linear plasmonic waveguide, and Eqs. (\ref{21})-(\ref%
{normalization}) reduces to the dispersion relation of linear plasmonic
modes, which have been discussed in Refs. \cite{OE22-24322,OL39-5909}.

\begin{figure}[ptb]
\centering
\vspace{0.0cm} \includegraphics[width=8cm]{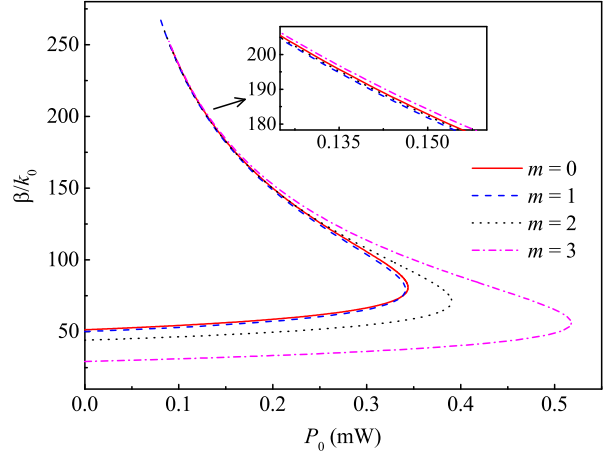} \vspace{-0.1cm}
\caption{Mode bifurcation curves with different orders. The
inset is the enlarged figure. The parameters are $f=30$ THz, $\protect\mu%
_{c}=0.3$ eV, $a=100$ nm, and $\protect\varepsilon_{1}=3$.}
\label{P0-beta}
\end{figure}

In what follows, we let $f=\omega/2\pi=30$ THz, $\mu_{c}=0.3$ eV, $a=100$
nm, and $\varepsilon_{1}=3$. Since the relaxation time of graphene ranges
from $0.01$ ps to $1$ ps \cite{PR535-101}, our parameters fulfill the
approximation conditions of $\hbar\omega\leq\mu_{c}$ and $\omega\tau\gg1$.
As shown in Fig. \ref{P0-beta}, graphene-coated dielectric nanowire supports
four orders of plasmonic modes, which
correspond to $m=0$, $1$, $2$, and $3$, respectively. Note when the input
power tends to zero, the propagation constants reduce to that of the
linear plasmonic modes\cite%
{OE22-24322}, where the nonlinear surface conductivity of graphene is neglected.
In other words, the linear plasmonic modes in Refs. \cite{OE22-24322}
are approximations of our nonlinear plasmonic modes in the limit of zero input power.

\begin{figure}[ptb]
\centering
\vspace{0.2cm} \includegraphics[width=8cm]{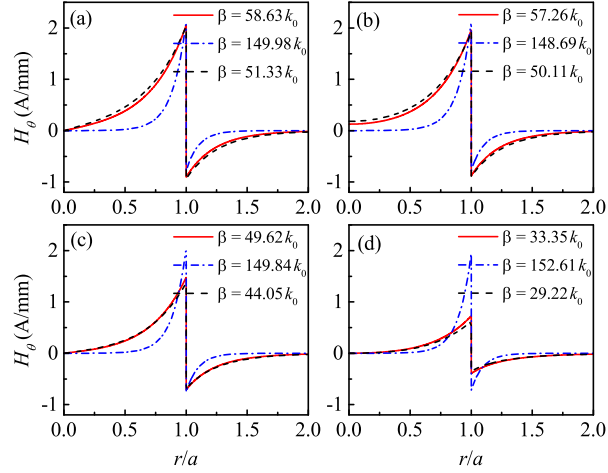} \vspace{-0.0cm}
\caption{The distribution of plasmonic modes in the radial direction at $\protect%
\theta=0$ for (a) $m=0$, (b) $m=1$, (c) $m=2$, and (d) $m=3$, respectively.
For comparison, the linear plasmonic modes are also plotted (black dashed
curves). The parameters are $f=30$ THz, $\protect\mu_{c}=0.3$ eV, $a=100$ nm,
$\protect\varepsilon_{1}=3$, and $P_{0}=0.2$ mW.}
\label{modes}
\end{figure}

Starting from the zero power point, as the input power increases along the
lower bifurcation branch, the field intensity at the graphene surface also
increases. According to Eqs. (\ref{sigma})-(\ref{sigma3}),
the surface conductivity of graphene decreases accordingly.
Thus, the propagation constant of
plasmonic mode increases with the enhancement of the
field confinement, as shown in Fig. \ref{modes}.
For comparison, we also plot the corresponding linear plasmonic modes.
Clearly, the propagation constants of these linear modes are equal to
that of the nonlinear plasmonic modes in the limit of zero input power.

In the lower bifurcation branch, the propagation constant increases monotonically
with the input power, when the input power is below a limitation value.
This limitation value is the maximum value of the allowed input power,
where the $m$th mode only exists when the input power is below its
corresponding limitation value. However, the field confinement can be
enhanced further if the input power decreases from the limitation value
along the upper bifurcation branch. Although the input power is decreased,
the field intensity at the graphene surface is increased which insures the
continuous growing of the propagation constant, as shown in Figs. \ref{P0-beta} and %
\ref{modes}.
Note although there are
intersection points when $P_0 = 0.339$ mW for the curves with $m=0$ and $m=1$,
and $P_0 =0.213$ mW for the curves with $m=0$ and $m=2$,
degenerate states do not exist since these
modes belong to different orders. Moreover, due to the high nonlinear
surface conductivity of graphene, the propagation constants for different orders
can be tuned by the
input power at the order of a few tenths of mW, which cannot been realized
by conventional nonlinear dielectric media \cite{PR535-101}.
Here we only show the four plasmonic modes which exhibit mode
bifurcation at low input power, although other modes may also be supported
by the graphene coated nanowire.

Comparing the mode bifurcation curves with different orders, the limitation values
for different plasmonic modes are not equal. At certain input powers, only the higher
order modes exist, and the lower order modes vanish.
This intriguing phenomenon is due to the self-action effect
of graphene. Since the surface conductivity of graphene is dependent on the
electric field that is parallel on the graphene surface, different
field intensities are required to support the plasmonic modes with different
orders, even at the same input power. Thus the nonlinear plasmonic waveguide may only
support the higher order modes at certain input powers.

\begin{figure}[ptb]
\centering
\vspace{0.1cm} \includegraphics[width=8cm]{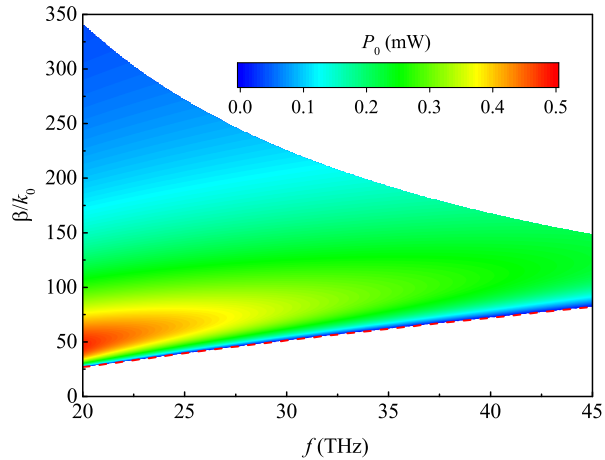} \vspace{-0.0cm}
\caption{Dispersion relation of the fundamental plasmonic mode with the dependence of the
input power. For comparison, the red dashed curve shows the dispersion
relation of the corresponding linear plasmonic mode. The parameters are $%
\protect\mu_{c}=0.3$ eV, $a=100$ nm, $\protect\varepsilon_{1}=3$, and $m=0$.}
\label{fig4}
\end{figure}

For the linear plasmonic modes in Refs. \cite{OL39-5909,OE22-24322}, the
dispersion relation between the propagation constant $\beta$ and frequency $f$
is a single curve, which is independent of the input power $P_{0}$. However, for
the nonlinear plasmonic modes, the dispersion curves at different input powers
form an energy
band, as shown in Fig. \ref{fig4}. For simplicity, we only consider the
fundamental mode with $m=0$, where the incident frequency is tuned between $f=20$
THz and $f=45$ THz to ensure the validity of the approximation conditions of
$\hbar\omega\leq\mu_{c}$ and $\omega\tau\gg1$.

Due to the nonlinearity of graphene, the plasmonic modes can only exist within a
certain power range. For lower frequencies, the propagation constant can be tuned
effectively by the input power. Whereas, the allowed band becomes narrow at
high frequencies with the decrease of the limitation value of the input
power.
Actually, as the input power
tends to zero, the dispersion relation reduces to that of the linear
plasmonic mode, as shown by the red dashed curve in Fig. \ref{fig4}.
Note the dispersion relation of the plasmonic mode would be different,
if the realistic loss of graphene is considered. The further research is beyond
the scope of this paper and we will show the relevant results else where.

In conclusion, considering the vector nature of plasmonic modes in two
dimensional waveguides, we derive the tensor form of nonlinear
surface conductivity of graphene. The plasmonic modes with
different orders are solved analytically in graphene-coated dielectric
nanowire, where the propagation constant of each mode can be tuned
effectively by the input power at the order of a few tenths of mW.
The lower and upper mode bifurcation branches are connected
at the limitation value of the input power. Moreover, due to the nonlinearity of
graphene, the dispersion curves of plasmonic modes at different input powers form an
energy band. Our work will provide important help to the research of other
graphene-based nonlinear waveguides, especially plasmonic waveguide
arrays and plasmonic lattices.

This work was sponsored by the National Natural Science Foundation of China
under Grants No. 61322501, No. 61574127, and No. 61275183, the Top-Notch Young
Talents Program of China, the Program for New Century Excellent Talents (NCET-12-0489)
in University, the Fundamental Research Funds for the Central Universities,
and the Innovation Joint Research Center for Cyber-Physical-Society System.


\begin{thebibliography}{99}
\bibitem{nature424-824} W. L. Barnes, A. Dereux, and T. W. Ebbesen, Nature
\textbf{424}, 824 (2003).

\bibitem{nphoton6-737} M. Kauranen and A. V. Zayats, Nat. Photon. \textbf{6}%
, 737 (2012).

\bibitem{SPJETPL32-512} V. M. Agranovich, V. S. Babichenko, and V. Ya.
Chernyak, Sov. Phys. JETP Lett. \textbf{32}, 512 (1980).

\bibitem{PRA28-1855} M. Y. Yu, Phys. Rev. A \textbf{28}, 1855 (1983).

\bibitem{OE17-21732} A. R. Davoyan, I. V. Shadrivov, and Y. S. Kivshar, Opt.
Exp. \textbf{17}, 21732 (2009).

\bibitem{PRA81-033850} A. Marini and D. V. Skryabin, Phys. Rev. A \textbf{81}%
, 033850 (2010).

\bibitem{JOSA72-1345} D. Sarid, R. T. Deck, and J. J. Fasanot, J. Opt. Soc.
Am. \textbf{72}, 1345 (1982).

\bibitem{OL9-235} G. I. Stegeman and C. T. Seaton, Opt. Lett. \textbf{9},
235 (1984).

\bibitem{JAP58-2460} J. Ariyasu, C. T. Seaton, G. I. Stegeman, A. A.
Maradudin, and A. F. Wallis, J. Appl. Phys. \textbf{58}, 2460 (1985).

\bibitem{OL32-674} E. Feigenbaum and M. Orenstein, Opt. Lett. \textbf{32},
674 (2007).

\bibitem{OE19-6616} A. Marini, D. V. Skryabin, and B. Malomed, Opt. Exp.
\textbf{19}, 6616 (2008).

\bibitem{OE16-21209} A. R. Davoyan, I. V. Shadrivov, and Y. S. Kivshar, Opt.
Exp. \textbf{16}, 21209 (2008).

\bibitem{PRA91-043815} A. Marini, S. Roy, Ajit Kumar, and F. Biancalana,
Phys. Rev. A. \textbf{91}, 043815 (2015).

\bibitem{PRL99-153901} Y. Liu, G. Bartal, D. A. Genov, and X. Zhang, Phys.
Rev. Lett. \textbf{99}, 153901 (2007).

\bibitem{PRL104-106803} F. Ye, D. Mihalache, B. Hu, and N. C. Panoiu, Phys.
Rev. Lett. \textbf{104}, 106802 (2010).

\bibitem{OE20-1945} J. Yan, L. Li, and J. Xiao, Opt. Exp. \textbf{20}, 1945
(2012).

\bibitem{OL37-2730} L. Wang, W. Cai, X. Zhang, and J. Xu, Opt. Lett. \textbf{%
37}, 2730 (2012).

\bibitem{PRB89-053406} Y. V. Bludov, D. A. Smirnova, Y. S. Kivshar, N. M. R.
Peres, and M. I. Vasilevskiy, Phys. Rev. B \textbf{89}, 035406 (2014).

\bibitem{EPL79-27002} S. A. Mikhailov, Europhys. Lett. \textbf{79}, 27002
(2007).

\bibitem{JP20-384204} S. A. Mikhailov and K. Ziegler, J. Phys. : Condens.
Mat. \textbf{20}, 384204 (2008).


\bibitem{PRL105-097401} E. Hendry, P. J. Hale, J. Moger, A. K. Savchenko and
S. A. Mikhailov, Phys. Rev. Lett. \textbf{105}, 097401 (2010)

\bibitem{PRB82-201402} K. L. Ishikawa, Phys. Rev. B \textbf{82}, 201402
(2010).

\bibitem{PR535-101} M. M. Glazov and S. D. Ganichev, Phys. Rep. \textbf{535}%
, 101 (2014).

\bibitem{LPR7-L7} M. L. Nesterov, J. B. Abad, A. Yu. Nikitin, F. J. G.
Vidal, and L. M. Moreno, Laser Photon. Rev. \textbf{7}, L7 (2013).

\bibitem{JPB46-155401} H. Dong, C. Conti, A. Marini, and F. Biancalana, J.
Phys. B \textbf{46}, 155401 (2013).

\bibitem{LPR8-291} D. A. Smirnova, I. V. Shadrivov, A. I. Smirnov, and Y. S.
Kivshar, Laser Photon. Rev. \textbf{8}, 291 (2014).

\bibitem{PRB91-045424} Y. V. Bludov, D. A. Smirnova, Y. S. Kivshar, N. M. R.
Peres, and M. I. Vasilevskiy, Phys. Rev. B \textbf{91}, 045424 (2015).

\bibitem{PRB90-125425} N. M. R. Peres, Yu. V. Bludov, Jaime E. Santos,
Antti-Pekka Jauho, and M. I. Vasilevskiy, Phys. Rev. B \textbf{90}, 125425
(2014).

\bibitem{nnano6-630} L. Ju, B. Geng, J. Horng, C. Girit, M. Martin, Z. Hao, H. A. Bechtel,
X. Liang, A. Zettl, Y. R. Shen, and F. Wang, Nat. Nanotech. \textbf{6}, 630
(2011).

\bibitem{SSC146-351} K. I. Bolotin, K. J. Sikes, Z. Jiang, M. Klima, G. Fudenberg, J. Hone,
P. Kim, H. L. Stormer, Solid State Commun. \textbf{146}, 351 (2008).

\bibitem{NO} R. W. Boyd, \textit{Nonlinear Optics}, 3rd ed. (Academic, San
Diego, 2008).

\bibitem{OE22-24322} Y. Gao, G. Ren, B. Zhu, H. Liu, Y. Lian, and S. Jian,
Opt. Exp. \textbf{22}, 24322 (2014).

\bibitem{OL39-5909} Y. Gao, G. Ren, B. Zhu, J. Wang, and S. Jian, Opt. Lett.
\textbf{39}, 5909 (2014).

%

\end{thebibliography}
\end{document}